# LOCAL ePOLITICS REPUTATION CASE STUDY

Jean-Marc.Seigneur@reputaction.com
*University of Geneva*
*7 route de Drize, Carouge, CH1227, Switzerland*

**ABSTRACT**

More and more people rely on Web information and with the advance of Web 2.0 technologies they can increasingly easily participate to the creation of this information. Country-level politicians could not ignore this trend and have started to use the Web to promote them or to demote their opponents. This paper presents how candidates to a French mayor local election and with less budget have engineered their Web campaign and online reputation. After presenting the settings of the local election, the Web tools used by the different candidates and the local journalists are detailed. These tools are evaluated from a security point of view and the legal issues that they have created are underlined.

## 1. INTRODUCTION

Web 2.0 technologies have increasingly facilitated the online publication of information by the users. The Web has become so pervasive that many politicians have adopted these Web technologies to promote their political parties, drive their election campaign and build their online political reputation. For example, during the 2007 French presidential elections, one of the main candidates proposed a Web-based participatory forum where users could submit ideas that other users could vote and comment. Similarly to the online tools presented in this paper, that participatory forum was not so secure, for example, it was possible to vote several times by creating different accounts. Another example is Barack Obama who has set up his own online social network with over 1.5 million friends and over 45 000 followers on Twitter [N08]. Even if the main candidates running for national political elections have an important budget, embracing a perfect online Web campaign is still very hard to achieve due to the distributed aspect of the Web where counter sites or defaming information can be propagated. In this paper, we investigate how candidates to a French mayor election and with less budget than national elections engineered their Web campaign. In Section 2, we describe the settings of the local election case study. These tools are then evaluated from a security point of view in Section 3. We draw our conclusion in Section 4.

## 2. DESCRIPTION OF THE LOCAL ELECTION AND ITS WEB TOOLS

We first present the general settings of the studied local election. Then, we detail the technical settings of the online tools used during this election.

## 2.1 Local Election General Settings

The studied local election happened in March 2008 during the French municipalities local elections. The studied municipality, namely Megève, one of the twelve Best of the Alps ski resorts, has around 3 900 voters. Two candidates ran for the elections to become the mayor of this municipality. The municipality politicians had been in place for twelve years for most of them. They had not to fight for their position six years ago at the time of the previous election because there was only one candidate. The status of the municipality has been quite worrisome because its population has decreased from more than 5 200 permanent people in the mid-eighties to around 4112 in 2007. This village is a well-known touristic place where the price of properties keep increasing. Due to too high French inheritance taxes, local people cannot afford to keep their family properties at time of inheritance. They have to sell their properties and then have to leave the village due to the high renting prices. The winning candidate obtained around 70% of the votes. Debates were quite fierce between the two candidates and as explained in the evaluation section below the online tools contributed a lot to increase the hostilities.
Concerning the legal settings, there are French laws (articles from L 47 to L 52-3 [CE]) related to online political elections that forbid the actions detailed in the following bullets. These bullets are used in Section 3 to evaluate the security and legal aspects of the different Web tools used during this election campaign:

- Defamation of a candidate, in our case via online information;
- Publication of new information by the candidate the day before the vote and the day of the vote, in our case online publication;
- Publication of surveys results that are not official or statistically correct without informing the readers that the results are not statistically correct;
- Collection of personal information by the candidates, for example, for political newsletters, without informing the French privacy protection agency called CNIL [CL] of the existence of this collection.

## 2.2 Local Election Technical Settings

The five main Web-based tools that were involved in the political debate are described below. The fifth one had been set up by us as part of a broader research project on the impact of online social networking tools on village communities. It is worth mentioning that it was the first time that the mayor local elections involved online tools in this municipality.

### 2.2.1 Candidate A's Web site

The official tool of the first candidate was a standard static Web site consisting of four HTML pages hosted by 1and1.com and mainly automatically generated by the 1and1 WYSIWYG Web site creator. The domain name was booked on the 24th of January 2008 and was the first to be publicly available. An online video of the candidate was available on the main page. The page views counts was managed by an image marker embedded by a third party traffic monitoring service. The cost of this Web site (excluding the cost of the knowledge and the two to five hours to set it up) is estimated at around 12 Euros, which corresponds to the price of the 1and1.com account for one year.

### 2.2.2 Candidate B's Blog

The second candidate set up a Typepad blog and bought the main domain name on the 9th of February 2008. The blog contained both static Web pages as it is possible to set up with Typepad and blog articles submitted from mid-February to the end of the elections in an increasing flow of information. People could subscribe to this blog and receive the articles through their blog reader. After each public meetings, YouTube videos of these meetings were embedded in articles or pages. No doubt that the WYSIWYG Typepad multi-authoring tools facilitated a lot the edition, the administration and the referencing of candidate B's blog. Although the comments were open until the day before the election, they did not use the feature to discuss negative comments. The advantage is that it seemed more dynamic through the daily submission of articles. The cost of that kind of blog (excluding the cost of the knowledge and the hours to edit the content) is estimated at around 45 Euros, which corresponds to the price of 3 months of Typepad use.

### 2.2.3 Candidate A Partisan's Blog

In October 2007, someone set up a blog to publish articles about what is going on in the village: the touristic events, the municipality projects... Based on the articles published by the authors of the blog, it seems clear that the authors of the blog were for candidate A who lost the elections. That blog is quite basic and has been set up on a French blog provider platform. Google Adsense advertisements are displayed throughout the blog pages and articles. The blog is still active. It was worth mentioning that blog because it contributed to the violence of the debates similarly to the main local newspaper site detailed below. Both tools are further discussed in the legal part of the evaluation section.

### 2.2.4 Main Local Newspaper's Web site

The main local newspaper provides a Web platform where its users can create an account and then leave comments on the online version of its articles published on its paper-based version. The comments are directly published and then manually moderated if someone reports an issue via an email. It is also possible to vote for an article on a 5-star scale or for a comment with a thumb-up or thumb-down without being logged in as depicted in Figure 1. The evaluation section below discusses where it failed to comply to the legal rules listed in Section 2.1.

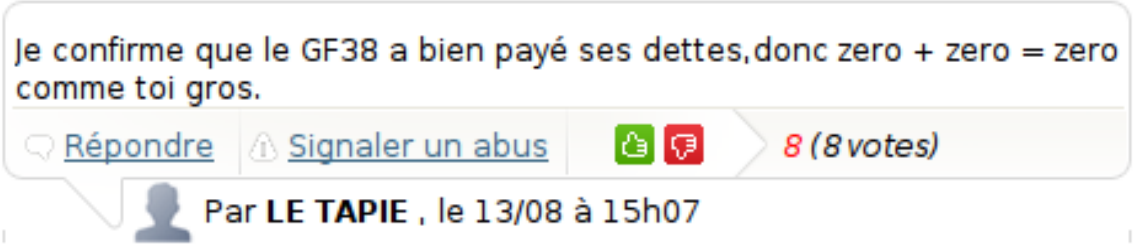

Figure 1. Main Local Newspaper Comment Rating

### 2.2.5 Our Local Online Social Networking Site

We designed and developed a Java-based online social networking Web site [MAG] dedicated to the municipality and its inhabitants to study the impact of such online social tools on such a municipality. The users used this site during the municipality campaign. From the beginning of August 2007 to the end of February 2008, 41 proposals to improve the village were made on our online social network contributed by 124 users connected between them as depicted in Figure 2. Each user is placed uniformly around a circle and a line, e.g. friendship or family relationship, represents a connection between a user and another user.

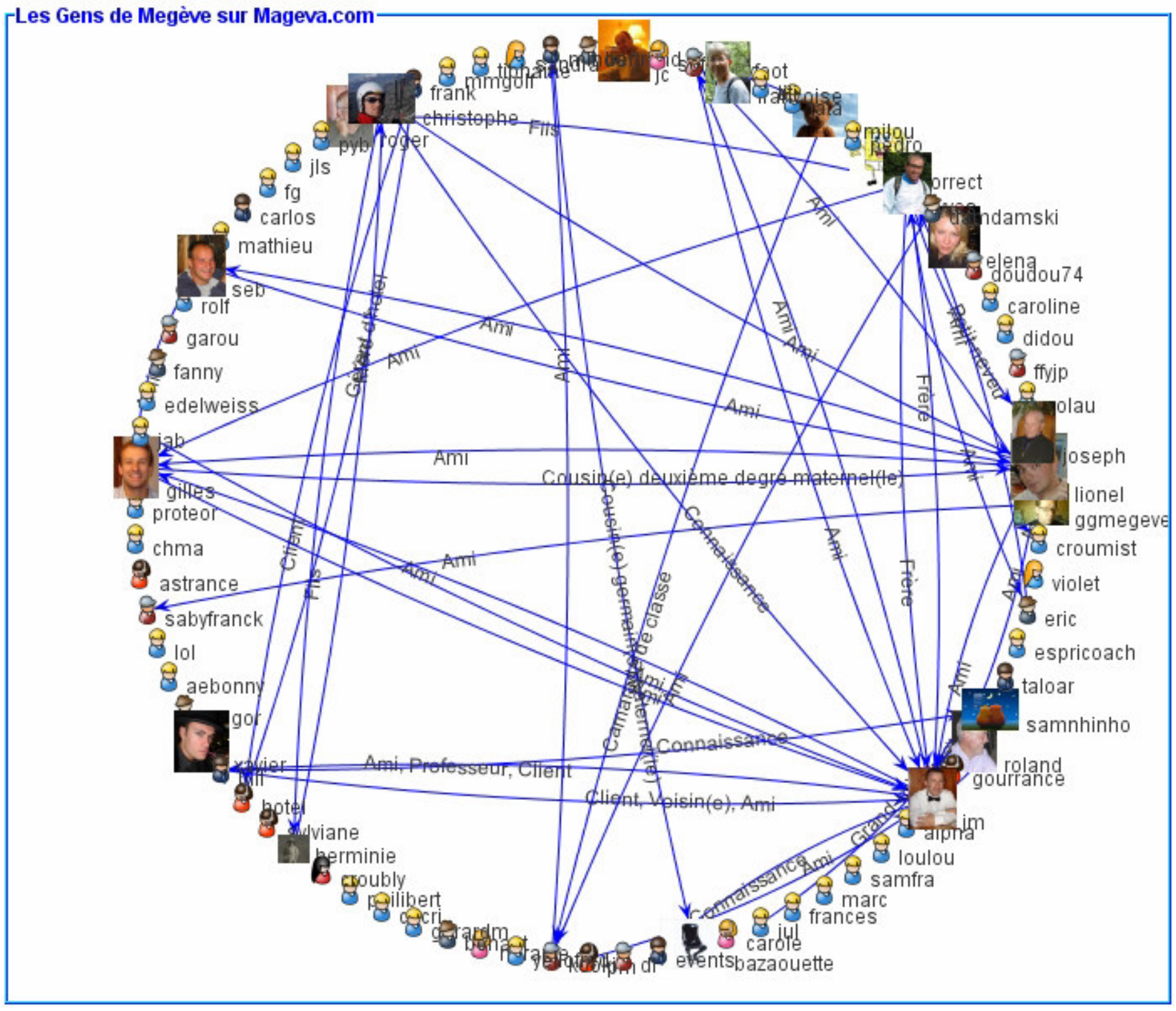

Figure 2. Local Online Social Network

The presentation layer of this site was based on JSP and JSF components from Apache MyFaces. The Web server used was Apache Tomcat. The Java objects were persisted in a MySQL database through an Hibernate persistence layer. The site was opened to the public in August 2007. The users could create an account linked to an email address, whose ownership is verified through an email challenge/response. At the beginning, they were only allowed to propose and comment new ideas to improve the village. Those propositions were listed in a forum section of the Web site. As they contributed, they earned participation points and the best participating users were promoted on a specific ranking page. There were also special users who had been coopted as real citizens of the village because they were known by the administrator of the site or other previously coopted users. Coopted users were allowed to vote for such or such proposition. To foster participation, the users could choose to remain anonymous when adding comments or propositions. The site is now based on Drupal and its list of functionalities has changed as well as its use, which is much lower than during the elections.

### 2.2.6 Overall Use of the Different Tools

Both candidates used a few other e-tools for their campaign. They submitted emailings through various means: the mailing list collected through their Web site, messages to Facebook groups (one candidate had set up a Facebook account), and allegedly even through unsolicited mass e-mailing as explained in the security evaluation below. To estimate that traffic, given that the counter on the main page of candidate's A Web site displayed 1 579 views on the 1st of March and that this Web site has 4 pages, around 6 000 pages may have been viewed on candidate's A Web site. Concerning candidate's B blog, the Typepad views tool counted 25 353 pages views over approximately the same period of time. Figure 3 shows the distribution of the number of page views per day. The first peak corresponds to when the local newspaper published its article describing the candidate B's team and project with a reference to the blog URL. The second peak corresponds to just after the election day and results. The number of views varied base on whether or not a blog article was posted on that day and if it contained a video of the new meetings or not. Thus, it seems that the blog approach attracted more page views than the static Web page approach.

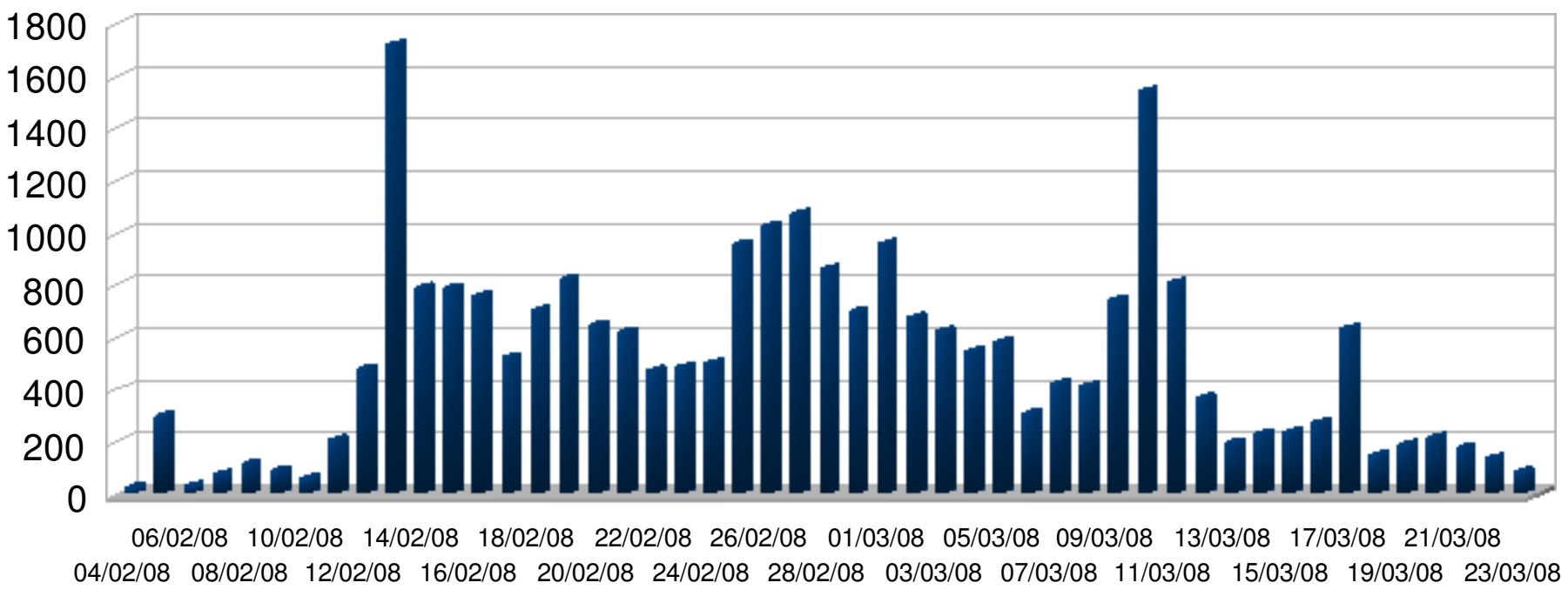

Figure 3. Candidate B's Blog Daily Number of Page Views

## 3. SECURITY DISCUSSION

In this section, we start by discussing the different legal aspects introduced in section 2.1:

- Declaration of personal data collection to the CNIL [CL]: Although it is mandatory to declare information collection and the declaration is quite easy through an online form, one of the candidates did not mention its CNIL declaration number on its Web site. In addition to this number, only one candidate respected the rule to clearly mention who was the editor of the information found on the Web pages. Another privacy issue happened because one of the candidates allegedly mass emailed to potential voters without having obtained their consent before submission of the emailing. Thus, anti-spam laws have not been respected. CNIL and editor information is present on the main local newspaper Web and our local social network. This point was less relevant to the partisan's blog.

- No publication of new information the day before the vote and the day of the vote: Both candidates seem to have respected this rule. The blog was not updated after the deadline and its comments features were disabled. The Web site of the other candidate did not change a lot anyway from the start to the end because it was not a blog with daily blog articles. The partisan's blog also respected this rule. The administrators of the main local newspaper platform had disabled the comments on the articles related to this election. We closed down our social network a few days before the elections in order to avoid any issue in this regard because we had not planned the functionality to forbid the users to publish information.
- No publication of surveys that are not statistically correct without mentioning it: Neither the candidates nor the partisan published surveys. As proposals to improve the village were voted and ranked on our online social network, we clearly mentioned that the ranking was not statistically correct and did not reflect the views of all the citizens. More detail on the number of people who participated to our social network is presented below. The main local newspaper did not clearly mention that the articles or comments ratings through the stars or the number of thumbs up and down were not representative and statistically correct. In addition, the votes on the comments were only protected by cookies and users could easily cheat, for example, just by clearing their cookies cache and voting again many times on the same comment. We contacted them to underline these issues but they had more important legal issues with their system as discussed below. Our online social network was protected against such automated votes thanks to captchas [AB03].
- Defamation: The main local newspaper system was involved in a defamation trial and we argue that it was mainly due to a bad design of their system. On the 24th of January, the local newspaper published an article describing the first candidate, his team and projects for the municipality. Comments started quickly to be published and attracted a lot of views. The day of the election, the article had been viewed 10 242 times and 406 comments had been added. Almost all comments have been voted and many of them have more than 100 votes. The pity is that these comments were not moderated before publication and many comments were quite harsh. Furthermore, at the beginning, the moderator who checks the comments had not really the time to keep checking these comments and did not think it was so important to moderate this thread. On the 11th of February, an article covered the other candidate profile, team and projects. Again this article attracted many views (7 378 before the election) and 298 comments. The comments becoming more and more known and harsher, the persons targeted by the comments started to complain to the newspaper. The moderator had to spend more time on checking in real time the flow of comments and eventually decided to close the comments submissions after one person sued for defamation. The last message from the moderator is copied in Figure 4 (to sum up, the text says that the comments feature has been disabled due to defamation trials, by the way, the author of the defamation was found by the police after investigations although he used a pseudonym). In contrast, our online social network experienced only one proposition to be moderated. It may be because the users felt that they were more easily recognized on our specific social network and more information was displayed to them regarding the risks that they were encountering. However, we had also to moderate the comments after they had been published, which is an issue since the comments still appear for some time on the Web site before being moderated.

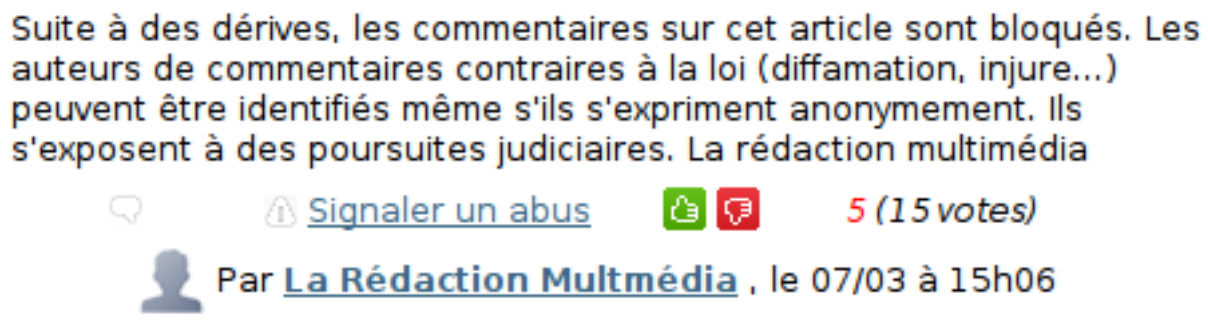

Figure 4. Final Newspaper Moderator Message

The situation is clearly unsatisfactory from a security point of view because many of the listed legal aspects are not fully respected. From a technical point of view, it seems clear that moderating before publication is a required feature. Another technical improvement is to use other protection means than simple cookies for vote protection. It should be costly and time consuming for users to cheat. From a user point of view, it is a pity that the newspaper did not inform and educate enough the users about the risks of publishing defaming information. Most of the users may have thought that they could not be found behind their pseudonyms. As said above, the person who published the defaming information involved in the trial has been found and sued in mid-July 2008. The newspaper should also have better explained that the results of the votes were not representative and not statistically correct. Another user comprehension issue related to security happened. Initially we added a captcha to the registration form of our online social network. Many aged users were not familiar with that kind of security mechanism and many users reported to us that they did not succeed to create an account due to these captchas. Our framework logged how many times users were calling the registration page and filling it with success. Before removing the captchas, only around one third of the loaded registration pages ended in a successful user registration with all fields and captcha correctly filled out. We eventually decided to remove this captcha on the registration page. From a security point of view, it is interesting to know how many times the users have chosen to use the anonymization feature on our online social network site when reacting to the 41 proposals published on the site. These 41 proposals created 81 comments and 29 of these comments were made with the anonymization option. Thus, around 35,8 % of the comments were anonymized, which may indicate that an anonymization feature may increase the level of participation.

# 4. CONCLUSION

Our case-study where local politicians tried to leverage the Web for their political success confirms Anderson and Cornfield's statement made in 2000: “there are growing signs that democratic politics will increasingly be conducted online” [AC02]. Park and Perry [PP07] found that the “use of campaign web sites is influential for giving money to political candidate and sending e-mails urging people to vote with and without reference to a particular candidate. However, the impact on voting is negligible”. We do not have the right data to apply their evaluation method. We can just report that the candidate who won the local election of our case-study was the candidate who used the more dynamic blogging platform. Local politicians are still not fully aware of the legal aspects that must be respected on the Web, especially regarding defamation. These issues also arise for the other actors, citizens and professionals such as the main local newspaper who did not put in place enough mechanisms to avoid defamation issues and who did not inform well-enough their users about their risks. These legal aspects will have to be addressed in the new open source or hosted technical solutions that may be proposed to more easily and more cheaply build participative and interactive political campaigns.

# REFERENCES

[AB03] Ahn, L.V., and Blum M., et al., 2003. captcha: using hard AI problems for security. *Proceedings of Eurocrypt.* Warsaw, Poland, pp. 139-149.

[AC02] Anderson, D. M., and Cornfield, M., 2002. *The Civic Web: Online Politics and Democratic Values.* Rowman & Littlefield.

[CE] Code électoral. http://www.legifrance.gouv.fr

[CL] CNIL. http://www.cnil.org

[MAG] http://www.mageva.com

[MEG] Megève. http://www.megeve.com

[N08] Nations, D., 2008. *How Barack Obama is using Web 2.0 to run for president.* About.com.

[PP07] Park, H. M., and Perry J. L., 2007. Do Campaign Web Sites Really Matter in Electoral Civic Engagement?: Empirical Evidence From the 2004 Post-Election Internet Tracking Survey. In Social Science Computer Review, Vol. 26, No. 2, pp 190-212.